\documentclass{article}
\usepackage{spconf,amsmath,graphicx}
\usepackage{tabularx,booktabs}
\usepackage{lipsum}
\usepackage{multirow}
\usepackage{booktabs}
\usepackage{cite}
\usepackage{url}

\title{Enabling Beam Search for Language Model-Based Text-to-Speech Synthesis}
\name{Zehai Tu$^1$, Guangyan Zhang$^1$, Yiting Lu$^1$, Adaeze Adigwe$^2$, Simon King$^2$, Yiwen Guo$^\dagger$\thanks{$^\dagger$Corresponding author}}
\address{$^1$LightSpeed Studios, Tencent, $^2$University of Edinburgh \\
zehaitu@global.tencent.com}

\renewenvironment{thebibliography}[1]{
  \begin{oldthebibliography}{#1}
    \setlength{\itemsep}{0.32em}
    \setlength{\parskip}{0em}
}
{
  \end{oldthebibliography}
}

\begin{document}
\ninept

\maketitle

\begin{abstract}

Tokenising continuous speech into sequences of discrete tokens and modelling them with language models (LMs) has led to significant success in text-to-speech (TTS) synthesis. Although these models can generate speech with high quality and naturalness, their synthesised samples can still suffer from artefacts, mispronunciation, word repeating, etc. In this paper, we argue these undesirable properties could partly be caused by the randomness of sampling-based strategies during the autoregressive decoding of LMs. Therefore, we look at maximisation-based decoding approaches and propose Temporal Repetition Aware Diverse Beam Search (TRAD-BS) to find the most probable sequences of the generated speech tokens. Experiments with two state-of-the-art LM-based TTS models demonstrate that our proposed maximisation-based decoding strategy generates speech with fewer mispronunciations and improved speaker consistency\footnote{See \url{tuzehai.github.io/trad-bs.github.io/} for samples}.

\end{abstract}


\section{Introduction}
\label{sec:intro}
TTS has seen remarkable advancements due to the adoption of LM-based paradigms in recent years\cite{wang2023neural, kharitonov2023speak, peng2024voicecraft, du2024cosyvoice, casanova2024xtts}. These LM-based TTS models are typically optimised to model sequences of speech tokens, which are quantised from continuous speech signals using pretrained neural speech tokenisers. Leveraging large speech datasets, these models can synthesise speech from a single prompt, effectively preserving the voice and emotion conveyed in the prompt. Furthermore, the quality, naturalness, and expressiveness of the synthesised speech are often impressively high.

However, generating satisfactory samples with LM-based TTS models can sometimes be challenging. Synthesised audios may exhibit artefacts, speaker inconsistency, mispronunciations, or even \textit{hallucinations}, such as word deletions or repetitions. These issues are more prevalent when synthesising long texts, texts with repetitive phrases, or when the speech prompt is highly emotional or acoustically complex. To address these challenges, various approaches have been proposed, including building more powerful models and using better datasets. Despite these efforts, the techniques for autoregressive decoding in LMs for TTS have received comparatively little attention.

Inherited from text generation, the autoregressive generation of tokens often relies on sampling-based strategies like \textit{top-p} \cite{holtzmancurious} or \textit{top-k} sampling in LM-based TTS models. The sampling-based decoding ensures the diversity of generation contribute to the expressive and human-like generation. However, speech has inherently lower information density and longer token sequences compared to text. For instance, an utterance containing ten words might require several hundred speech tokens. Consequently, the randomness introduced by sampling-based decoding could lead to greater instability in speech token generation, often resulting in artefacts and mispronunciations.

Motivate by this, a maximisation-based decoding strategy is explored in this paper for LM-based TTS. We first examine beam search (BS), arguably the most popular maximisation-based approach, and demonstrate its failure in the autoregressive generation of speech tokens. To address these pitfalls, we propose Temporal Repetition Aware Diverse Beam Search (TRAD-BS), designed to generate the most probable speech token beams. We validate our approach using two state-of-the-art LM-based TTS models across several datasets. Both objective and subjective evaluations indicate that TRAD-BS outperforms sampling-based decoding strategies.


\begin{figure*}[ht!]
    \centering
    \includegraphics[width=0.75\textwidth]{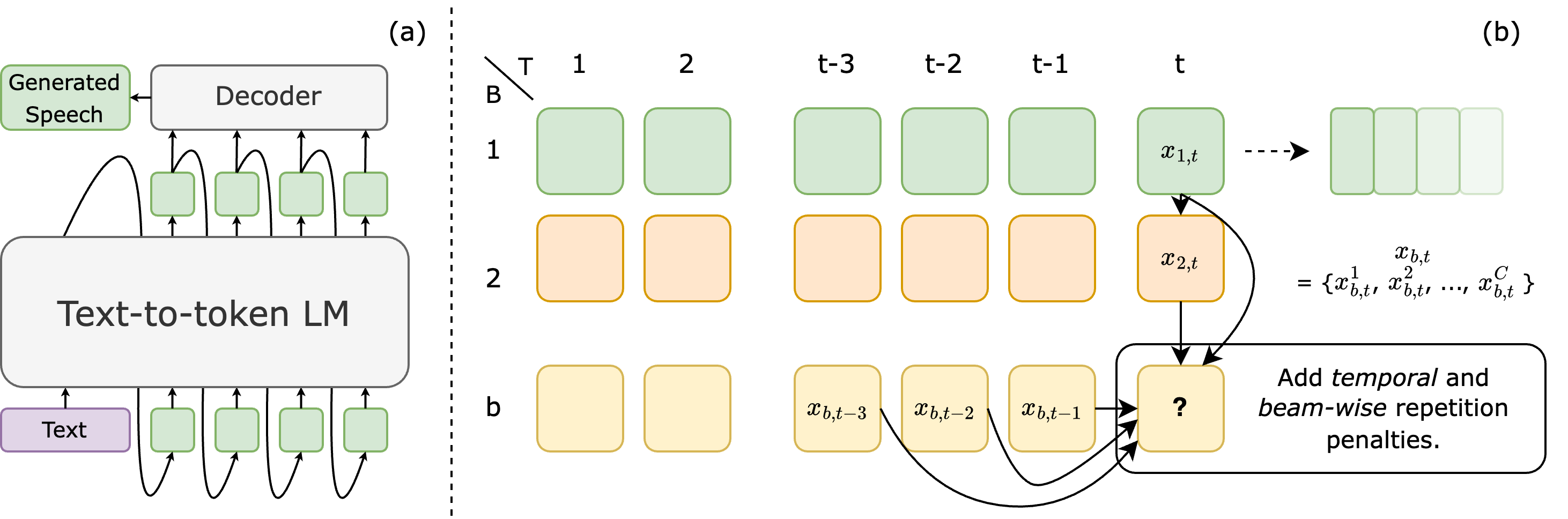}
    \caption{(a) A general inference paradigm of LM-based TTS. Speech tokens are generated autoregressively before converted into speech. (b) TRAD-BS operates through both decoding steps and beams, adding temporal and beam-wise penalties to repeated tokens. This can also be applied to the case where multiple tokens need to be decoded at a single step.}
    \label{fig:method}
\end{figure*}

\section{Background}
\label{sec:bg}

\subsection{LM-based TTS}
In recent years, LMs trained on sequences of speech tokens have garnered significant attention due to their high-quality generation and zero-shot capabilities. This approach was initially introduced in the context of textless NLP \cite{lakhotia2021generative}, aimed at building dialogue systems using only audio data. The concept was further developed with the introduction of AudioLM \cite{borsos2023audiolm}, which demonstrated strong performance in speech continuation by modelling a combination of acoustic and semantic speech tokens. LM-based approaches were subsequently extended to zero-shot TTS tasks, as seen in models like VALL-E \cite{wang2023neural} and Spear-TTS \cite{kharitonov2023speak}. By treating zero-shot TTS as a task of transcript-conditioned speech continuation, these models significantly outperformed previous methods. Another notable LM-based zero-shot TTS model, XTTS \cite{casanova2024xtts}, utilises speaker embeddings as conditioning factors during synthesis. More recently, there has been a growing interest in applying LM-based approaches to TTS. For example, VoiceCraft \cite{peng2024voicecraft} adapted LMs for speech editing, while natural language descriptions have been explored as guidance for TTS \cite{lyth2024natural}. Additionally, studies have shown that LM-based TTS systems can greatly benefit from an increasing amount of training data \cite{anastassiou2024seed, du2024cosyvoice, lajszczak2024base}.

A common paradigm in LM-based TTS involves modelling sequences of paired text and speech tokens using a text-to-token LM. The choice of speech tokeniser is crucial to the overall performance of the TTS system. Many works, such as \cite{wang2023neural, defossezhigh}, utilise neural Codec models \cite{defossezhigh, zeghidour2021soundstream} to convert speech signals into acoustic tokens, which typically capture the fine structure of speech signals. These neural Codec models often employ multiple codebooks for tokenisation using the residual vector quantisation (RVQ) technique, resulting in a speech representation for a single frame that consists of multiple tokens. At the same time, semantic tokens have also proven effective in achieving high-quality speech synthesis \cite{lajszczak2024base, du2024cosyvoice}. These semantic tokens can be quantised using self-supervised learning models \cite{hsu2021hubert,chen2022wavlm} or automatic speech recognition models \cite{du2024cosyvoice}.

The general inference process for LM-based TTS is illustrated in Figure~\ref{fig:method}(a). The text-to-token LM generates speech tokens in an autoregressive manner, which are then converted into speech signals using a decoder. Neural Codec decoders are commonly used for acoustic tokens, while diffusion-based generative models are particularly effective in generating high-quality audio from semantic tokens.

\subsection{Decoding strategies}
The decoding strategies used in autoregressive generation play a critical role in determining the performance of language models (LMs). One widely adopted strategy is \textit{top-k} sampling, where the next token is sampled from the top \textit{k} most likely options. This approach was instrumental in enabling GPT-2 to achieve impressive generation quality \cite{radford2019language}. However, selecting the appropriate \textit{k} value can be challenging. A low \textit{k} might produce outputs that are too predictable, resembling maximisation methods, while a high \textit{k} can introduce excessive variance, leading to observable incoherence in the generated text. To address these issues, \textit{top-p} sampling was introduced \cite{holtzmancurious}. Instead of sampling from a fixed number of tokens, \textit{top-p} sampling selects from the smallest set of tokens whose cumulative probability exceeds a threshold \textit{p}. This method has become widely used in LMs due to its ability to balance diversity and coherence in generated outputs. On the other hand, maximisation-based decoding approaches, such as beam search (BS), often fail to produce ideal text generation results. While BS can generate sequences that are more deterministic, the outputs typically lack diversity and can be degenerate. Early efforts to enhance BS by encouraging the generation of diverse beams include works like those in \cite{li2016simple, vijayakumar2016diverse, kulikov2019importance}.


\section{Method}
\label{sec:method}
In this section, we investigate maximisation-based decoding strategies with BS, a widely used method for finding sequences with maximal likelihood. First, we briefly formulate the BS algorithm. Then, we demonstrate why BS cannot be directly applied to LM-based TTS. Finally, we address the limitations of BS and introduce our proposed decoding method to overcome these challenges.

\subsection{BS formulation}
BS is a heuristic search algorithm designed to find the most probable sequence during decoding. At each time step $t$, BS retains the top-$B$ highest-scoring candidates, where $B$ is known as the beam width, i.e., the number of beams. Let the set of $B$ candidates at time step $t-1$ be denoted as:
\begin{equation}
X_{t-1} = \{\mathbf{x}_{1, t-1}, \mathbf{x}_{2, t-1}, \ldots, \mathbf{x}_{B, t-1}\},
\end{equation}
where each candidate is represented as a sequence of tokens:
\begin{equation}
\mathbf{x}_{b, t-1} = \{x_{b, 1}, x_{b, 2}, \ldots, x_{b, t-1}\}.
\end{equation}
At each time step, BS examines all possible token extensions of all beams given by the set $\mathcal{X}=X_{t-1} \times \mathcal{V}$, and keep the $B$ most likely sequences, denoted as:
\begin{equation}
X_{t}=\underset{\mathbf{x}_{1,t}, \ldots, \mathbf{x}_{B,t} \in \mathcal{Y}_t}{\operatorname{argmax}} \sum_{t=1} \log p\left(x_{b,t}\right),
\end{equation}
where $\mathcal{V}$ represents the vocabulary of tokens. In many neural Codec LMs, multiple codebooks are used to quantize a speech frame at a single time step, meaning $x_{b, t}$ can consists of $C$ tokens corresponding to the number of codebooks:
\begin{equation}
x_{b, t} = \{x^1_{b, t}, x^2_{b, t}, \ldots, x^C_{b, t} \}.
\end{equation}
In this case, the subscript for codebook $c$ will be omitted for simplicity for the rest of this paper.

\subsection{Pitfalls of BS in LM-based TTS}
Although BS s designed to identify the sequence with the highest probability in autoregressive decoding, it encounters significant issues when applied to LM-based TTS. These issues manifest in two main ways. 

\begin{figure}[t]
    \centering
    \includegraphics[width=0.8\linewidth]{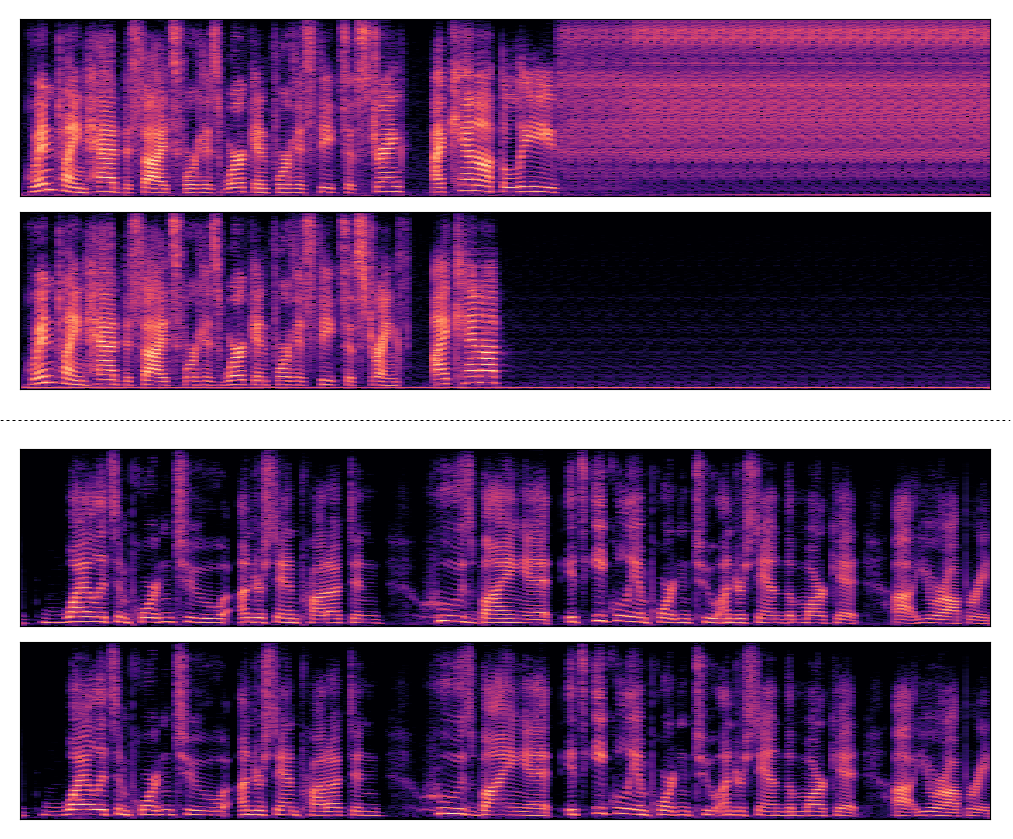}
    \caption{Examples of the temporal collapse (upper) and the beam-wise diversity collapse (lower) of BS in LM-based TTS. }
    \label{fig:collapse}
\end{figure}

First, BS can suffer from temporal collapse, where the generated sequences consist of repeated tokens or very short token combinations. This issue results in generated speech that is stuck in silence or constant artefacts. As illustrated in the upper part of Figure~\ref{fig:collapse}, the initial segments exhibiting recognisable speech patterns correspond to the speech prompts, while the following generated segments consist of either noise or silence. This problem can occur with any type of greedy decoding, as well as with sampling-based strategies when the temperature is set too low, reducing randomness and leading to minimal variability in sampling.

Second, BS often experiences beam-wise diversity collapse, where the token sequences across different beams exhibit minimal differences. This results in nearly identical generated speech from different beams. As shown in the lower part of Figure~\ref{fig:collapse}, the spectrograms of speech generated from two distinct beams are indistinguishable, highlighting the lack of diversity in the generated outputs.

\subsection{TRAD-BS}
To tackle the two types of collapses, we propose TRAD-BS, which penalise token repetition both across decoding steps and among different beams, as shown in Figure~\ref{fig:method}(b). It is worth noting that temporal repetition aware penalty has been used  in sampling-based decoding to avoid token repetition\cite{chen2024vall}. Beam-wise repetition penalty for beam diversity has shown its success in text generation \cite{vijayakumar2016diverse}, and firstly applied in TTS to the best of our knowledge.

Denoting $S_{t} = \{x_{b, t-l}, x_{b, t-l+1}, \ldots, x_{b, t}\}$ as the set of tokens in a window of length $l$ of beam $b$, and $S_{b} = \{x_{1, t}, x_{2, t}, \ldots, x_{b-1, t}\}$ as the set of tokens in the previous beams at step $t$, the log-probability of token $x_{b, t}$ is modified as follows:
\begin{equation}
\log p'(x_{b,t}) = \begin{cases} 
\alpha \log p(x_{b,t}), & \text{if } x_{b,t} \in S_{t}\setminus S_{b} \\
\beta \log p(x_{b,t}), & \text{if }x_{b,t} \in S_{b}\setminus S_{t} \\
\alpha\beta \log p(x_{b,t}), & \text{if } x_{b,t}, \in S_{t} \cap S_{b} \\
\log p(x_{b,t}), & \text{otherwise}
\end{cases}
\end{equation}

Here, $\alpha$ and $\beta$ are the temporal and beam-wise repetition penalty coefficient, respectively. Since log-probabilities are always less than 0, the coefficients $\alpha$ and $\beta$ are set to values greater than 1 to penalise repetition. The scoring of each beam at step $t$ is calculated with $\log p'(x_{b,t})$, while the log-probabilities of previous steps remain the unchanged. After decoding is complete, the beams are re-ranked based on the sum of the original $\log p(x_{t})$. In addition, each beam remains fixed through out the decoding process and not be pruned, otherwise the beam-wise diversity collapse can still happen.

\section{Experiments}
\label{sec:exp}

\subsection{Baselines}
For our experiments, we utilised two recently released state-of-the-art LM-based zero-shot TTS models: VoiceCraft\cite{peng2024voicecraft} and CosyVoice\cite{du2024cosyvoice}. They achieved impressive performance on English and Chinese, respectively, and are briefly introduced below.

\subsubsection{VoiceCraft}
VoiceCraft is a neural Codec LM system originally designed for speech editing but also performs exceptionally well in zero-shot TTS. It employs a transformer decoder to model acoustic tokens, which are encoded by the Encodec \cite{defossezhigh} encoder. The speech is synthesized from the generated tokens using the Encodec decoder. Due to the Residual Vector Quantization (RVQ) mechanism employed by Encodec, multiple codebooks are required for tokenization, resulting in the prediction of multiple tokens in parallel at each decoding step. During inference, the transformer decoder takes a prompt speech with its transcript, along with the transcript to be synthesized. The \textit{top-p} sampling strategy is employed for autoregressive token generation.

In our experiments, we used the released 830M VoiceCraft transformer model alongside the corresponding 4-codebook Encodec\footnote{huggingface.co/pyp1/VoiceCraft/tree/main}. The models were trained on the GigaSpeech\cite{chen21o_interspeech} training set, which comprises 9,000 hours of audiobooks, podcasts, and YouTube audio. VoiceCraft has demonstrated the ability to surpass several recent zero-shot models on both objective and subjective metrics when evaluated on the LibriTTS\cite{zen19_interspeech} and GigaSpeech test sets.

\subsubsection{CosyVoice}
CosyVoice is a recently proposed LM-based system for zero-shot speech synthesis, designed to model supervised semantic tokens. Specifically, these speech tokens are obtained by quantizing the hidden representations from an encoder layer of a powerful speech recognition model. These tokens are strongly correlated with textual information and offer advantages over unsupervised tokens in terms of content consistency and speaker similarity. After the tokens are generated by the LM transformer, a conditional flow matching model \cite{mehta2024matcha} is used to synthesise mel spectrograms, which are then converted into waveform audio via a vocoder. In addition to the transcripts and prompt speech, the speaker embedding from the prompt speech is also used during inference. The \textit{top-k} sampling strategy is employed for the autoregressive token generation process.

For our experiments, we used the released CosyVoice-300M\footnote{modelscope.cn/models/iic/CosyVoice-300M} model . It was trained on a multilingual dataset containing over 170,000 hours of speech, with 130,000 hours in Chinese. Objective evaluation results on AISHELL-3 \cite{shi21c_interspeech} indicate that CosyVoice achieves oracle-level performance in speech recognition and speaker similarity.

\subsection{Datasets}
For the English zero-shot TTS evaluation with VoiceCraft, we used a subset of the GigaSpeech test set and the SeedTTS English test set\cite{anastassiou2024seed}. The GigaSpeech evaluation set includes 150 samples from podcasts and YouTube audio. The voice prompt segments were kept around 3 seconds and were cut only between words. The SeedTTS English test set comprises over 1,000 samples from the Common Voice dataset\cite{ardila2020common}.

Similarly, two datasets were used, for the Chinese zero-shot TTS evaluation with CosyVoice. The first consists of 180 expressive utterances from an internal database, produced by professional voice actors and covering a range of emotions including happiness, sadness, pain, peacefulness, and others. The second test set is the SeedTTS Chinese test set, which includes over 2,000 samples from the DidiSpeech dataset \cite{guo2021didispeech}.
\begin{table}[t!]
\centering
\caption{\scriptsize Word/Character Error Rate (WER/CER) \& Speaker Similarity (SS) comparison between the Baseline sampling-based decoding and TRAD-BS decoding. VoiceCraft and CosyVoice models are evaluated respectively on two English (GigaSpeech, SeedTTS-EN) and two Chinese (Internal, SeedTTS-CN) evaluation sets. First, Best, and Mean represent the results of 5 beams or samples from TRAD-BS and baseline.}
\resizebox{\linewidth}{!}{
\begin{tabular}{ll|lll|lll}
\toprule
\multicolumn{2}{c|}{\multirow{2}{*}{VoiceCraft}} & \multicolumn{3}{c|}{WER $\downarrow$} & \multicolumn{3}{c}{SS$\uparrow$} \\
\multicolumn{2}{c|}{} & \multicolumn{1}{c}{First} & \multicolumn{1}{c}{Mean} & \multicolumn{1}{c|}{Best} & \multicolumn{1}{c}{First} & \multicolumn{1}{c}{Mean} & \multicolumn{1}{c}{Best} \\ \midrule
\multirow{2}{*}{GigaSpeech} & Baseline & 5.714 & 5.616 & 1.825 & 0.521 & 0.526 & 0.583 \\
 & TRAD-BS & \textbf{3.732} & \textbf{3.604} & \textbf{1.748} & \textbf{0.553} & \textbf{0.561} & \textbf{0.603} \\ \midrule
\multirow{2}{*}{SeedTTS-EN} & Baseline & 2.117 & 3.736 & 0.335 & 0.456 & 0.461 & 0.534 \\
 & TRAD-BS & \textbf{1.527} & \textbf{1.418} & \textbf{0.300} & \textbf{0.498} & \textbf{0.504} & \textbf{0.551} \\ \midrule \midrule
\multicolumn{2}{c|}{\multirow{2}{*}{CosyVoice}} & \multicolumn{3}{c|}{CER$\downarrow$} & \multicolumn{3}{c}{SS$\uparrow$} \\
\multicolumn{2}{c|}{} & \multicolumn{1}{c}{First} & \multicolumn{1}{c}{Mean} & \multicolumn{1}{c|}{Best} & \multicolumn{1}{c}{First} & \multicolumn{1}{c}{Mean} & \multicolumn{1}{c}{Best} \\ \midrule
\multirow{2}{*}{Internal} & Baseline & 7.999 & 7.987 & 4.264 & \textbf{0.840} & \textbf{0.840} & \textbf{0.858} \\
 & TRAD-BS & \textbf{5.174} & \textbf{4.971} & \textbf{3.135} & 0.838 & 0.838 & 0.855 \\ \midrule
\multirow{2}{*}{SeedTTS-CN} & Baseline & 3.833 & 3.860 & 1.191 & \textbf{0.847} & \textbf{0.847} & \textbf{0.866} \\
 & TRAD-BS & \textbf{2.088} & \textbf{2.146} & \textbf{1.164} & 0.825 & 0.827 & 0.844 \\ \bottomrule
\end{tabular}}
\label{table1}
\end{table}

\begin{table}[t]
\centering
\caption{\scriptsize Word/Character Error Rate (WER/CER) \& Speaker Similarity (SS) comparison across different beams from TRAD-BS.}
\resizebox{\linewidth}{!}{
\begin{tabular}{c|cccc|cccc}
\toprule
\multirow{3}{*}{Beam} & \multicolumn{4}{c|}{VoiceCraft} & \multicolumn{4}{c}{CosyVoice} \\
 & \multicolumn{2}{c}{GigaSpeech} & \multicolumn{2}{c|}{SeedTTS-EN} & \multicolumn{2}{c}{Internal} & \multicolumn{2}{c}{SeedTTS-CN} \\
 & WER$\downarrow$ & SS$\uparrow$ & WER$\downarrow$ & SS$\uparrow$ & CER$\downarrow$ & SS$\uparrow$ & CER$\downarrow$ & SS$\uparrow$ \\ \midrule
1 & \multicolumn{1}{c|}{3.732} & \multicolumn{1}{c|}{0.553} & \multicolumn{1}{c|}{1.527} & 0.498 & \multicolumn{1}{c|}{5.174} & \multicolumn{1}{c|}{0.838} & \multicolumn{1}{c|}{\textbf{2.088}} & 0.825 \\
2 & \multicolumn{1}{c|}{3.962} & \multicolumn{1}{c|}{0.558} & \multicolumn{1}{c|}{1.306} & 0.503 & \multicolumn{1}{c|}{\textbf{4.582}} & \multicolumn{1}{c|}{0.838} & \multicolumn{1}{c|}{2.094} & 0.826 \\
3 & \multicolumn{1}{c|}{3.688} & \multicolumn{1}{c|}{\textbf{0.567}} & \multicolumn{1}{c|}{\textbf{1.252}} & 0.506 & \multicolumn{1}{c|}{5.125} & \multicolumn{1}{c|}{\textbf{0.839}} & \multicolumn{1}{c|}{2.172} & 0.828 \\
4 & \multicolumn{1}{c|}{\textbf{3.216}} & \multicolumn{1}{c|}{0.561} & \multicolumn{1}{c|}{1.423} & \textbf{0.508} & \multicolumn{1}{c|}{4.906} & \multicolumn{1}{c|}{0.839} & \multicolumn{1}{c|}{2.208} & 0.829 \\
5 & \multicolumn{1}{c|}{3.422} & \multicolumn{1}{c|}{0.564} & \multicolumn{1}{c|}{1.581} & 0.506 & \multicolumn{1}{c|}{5.067} & \multicolumn{1}{c|}{0.838} & \multicolumn{1}{c|}{2.169} & \textbf{0.829} \\ \bottomrule
\end{tabular}}
\end{table}

\subsection{Evaluation}
For the objective evaluation, word error rates (WER) of the synthesised speech are measured against the target transcripts. Additionally, speaker similarities (SS) between the prompt speech and the generated speech are assessed by calculating the cosine distances between their speaker embeddings. English WER and SS are measured using the Whisper-medium model\cite{radford2023robust} and the pyannote.audio SinceNet-based XVector\footnote{https://huggingface.co/pyannote/embedding} model\cite{snyder2018x}, respectively. For the Chinese evaluations, the Paraformer  \cite{gao22b_interspeech} and CAM++ models \cite{wang23ha_interspeech} are used to obtain character error rates (CER) and SS\footnote{https://github.com/modelscope/FunASR}.

For the subjective evaluation, pairs of speech generated by the baseline models and with TRAD-BS were presented to listeners, who were asked to choose their preferred samples. In the VoiceCraft experiment, 25 listeners evaluated 40 sample pairs from the GigaSpeech test set. For the CosyVoice experiment, 25 listeners participated in listening tests of 36 sample pairs from the internal database.

\subsection{Setup}
In all experiments, a beam width of 5 was used, along with a window length of 50 for checking temporal repetition. For VoiceCraft, the temporal and beam-wise repetition penalty coefficients were set to 10 and 3, respectively. For CosyVoice, these coefficients were set to 15 and 10. For the objective evaluation of the baselines, we conducted sampling-based decoding using the released configurations of the TTS models. This was done 5 times with different random seeds. The objective metrics compared the first sample from the baseline and the first beam from the TRAD-BS decoding, as well as the mean and best results among the 5 samples and 5 beams.

\begin{table}[t]
\centering
\caption{\scriptsize Subjective preference between the Baseline sampling-based decoding and TRAD-BS decoding.}
\resizebox{0.5\linewidth}{!}{
\begin{tabular}{lcc}
\toprule
 & Baseline & TRAD-BS \\ \midrule
VoiceCraft & 28.05\% & 71.95\% \\ \midrule
CosyVoice & 37.70\% & 62.30\% \\ \bottomrule
\end{tabular}}
\label{subjective}
\end{table}

\section{Results}
\label{sec:results}
The objective evaluation results are presented in Table~\ref{table1}. The GigaSpeech utterances, originating from podcasts and YouTube audios, exhibit greater diversity in emotions, speaking rates, and acoustic environments compared to the SeedTTS-EN test set. Consequently, synthesised speech from GigaSpeech is more likely to produce artefacts, mispronunciations, or even complete distortion, leading to higher WERs. Since VoiceCraft is trained with the GigaSpeech dataset, it generally shows better speaker similarity (SS) on the GigaSpeech test set. For both test sets, TRAD-BS demonstrates significant improvements over \textit{top-p} sampling in terms of WER and SS for both the first and mean scores across five beams or samples. This suggests that the proposed maximisation decoding strategy generally produces samples with fewer mistakes and better speaker consistency. Notably, the best sample out of the 5 candidates often shows substantial improvement, and TRAD-BS consistently outperforms the baseline decoding strategy. This indicates that LM-based TTS models benefit from post re-ranking using beams from TRAD-BS.

For the CosyVoice Chinese zero-shot TTS evaluation, the Internal dataset is significantly more expressive than the read speech from SeedTTS-CN, leading to higher ASR recognition errors. TRAD-BS consistently outperforms the \textit{top-k} sampling decoding in terms of CER. However, the SS of the baseline samples is higher than that of TRAD-BS generated samples. Similar trends are observed in SeedTTS \cite{anastassiou2024seed}, where SS for synthesised samples is notably higher than for reference human speech. Perception differences are minimal when SS values are sufficiently high, e.g., above 0.7.

Additionally, we analysed performance across different beams in the four evaluation sets. It was observed that the first beams, which have the highest overall probabilities, do not necessarily achieve the best W(C)ER and SS scores. Despite the absence of a clear pattern for the best overall performance, the results generally show that maximisation-based BS decoding can outperform baseline sampling-based decoding. This is likely because the top beams in maximisation-based BS decoding have higher probabilities than sampling-based samples, and the marginal probability differences among them are insufficient to create significant WER or SS differences.

The subjective evaluation results, shown in Table~\ref{subjective}, indicate that TRAD-BS offers a clear advantage over sampling-based decoding strategies. The advantage is less pronounced for CosyVoice, which is trained with significantly more data than VoiceCraft, and the GigaSpeech samples have more complex acoustic environments than the internal test set. In general, TRAD-BS generated speech samples exhibit better quality, fewer mispronunciations, and improved speaker consistency. However, many samples generated by sampling-based methods are more expressive due to \textit{sampling-produced} prosody variations, which can also sometimes lead to unnatural tones and pauses.

\section{Conclusions}
\label{sec:conclusions}

In this paper, we enabled the beam search for LM-based TTS and proposed TRAD-BS. Instead of sampling the speech token at each autoregressive decoding step, TRAD-BS was designed to find the probable sequences incorporating temporal and beam-wise token repetition penalties. Both objective and subjective evaluations with two state-of-the-art TTS model demonstrated the advantage of TRAD-BS over sampling-based decoding methods.

\bibliographystyle{IEEEbib}
\bibliography{refs}

\end{document}